\documentclass[pdflatex,sn-nature]{sn-jnl}

\usepackage{graphicx}%
\usepackage{multirow}%
\usepackage{amsmath,amssymb,amsfonts}%
\usepackage{amsthm}%
\usepackage{mathrsfs}%
\usepackage[title]{appendix}%
\usepackage{xcolor}%
\usepackage{textcomp}%
\usepackage{manyfoot}%
\usepackage{booktabs}%
\usepackage{algorithm}%
\usepackage{algorithmicx}%
\usepackage{algpseudocode}%
\usepackage{listings}%

\usepackage[english]{babel}
\usepackage{mathtools}
\usepackage{amsmath}
\usepackage{amsthm}
\usepackage{amsfonts}
\usepackage{float}
\usepackage{graphicx}
\usepackage{caption}
\usepackage{subcaption}
\usepackage{physics}
\usepackage{url}
\usepackage{epstopdf}
\usepackage{lipsum}
\usepackage[title]{appendix}
\usepackage{titlesec}
\usepackage{gensymb}
\usepackage[separate-uncertainty = true]{siunitx}
\usepackage{colortbl}
\usepackage{wrapfig}
\usepackage{setspace}
\usepackage{amssymb}
\usepackage[ampersand]{easylist}
\usepackage{array}
\usepackage{cancel}
\usepackage[T1]{fontenc}
\usepackage[utf8]{inputenc}
\usepackage{soul}
\usepackage{xcolor}
\sethlcolor{yellow}
\usepackage[capitalise,sort&compress]{cleveref}
\crefname{equation}{Eq.}{Eqs.}
\crefname{figure}{Fig.}{Figs.}
\crefname{tabular}{Table}{Tables}
\crefname{algocf}{Algorithm}{Algorithms}
\crefname{remark}{Remark}{Remarks}
\crefname{theorem}{Theorem}{Theorems}

\Crefname{equation}{Equation}{Equations}
\Crefname{figure}{Figure}{Figures}
\Crefname{tabular}{Table}{Tables}
\Crefname{algocf}{Algorithm}{Algorithms}
\Crefname{remark}{Remark}{Remarks}
\Crefname{theorem}{Theorem}{Theorems}

\graphicspath{ {./Images/}}

\theoremstyle{thmstyleone}%
\theoremstyle{thmstyletwo}%

\theoremstyle{thmstylethree}%

\begin{document}

\title[Article Title]{\begin{centering}An Internal Digital Image Correlation Technique for\\ High-Strain Rate Dynamic Experiments\end{centering}}


\author*[1]{\fnm{Barry P.} \sur{Lawlor}}\email{blawlor@caltech.edu}

\author[2]{\fnm{Vatsa} \sur{Gandhi}}\email{vbg24@cam.ac.uk}

\author[1]{\fnm{Guruswami} \sur{Ravichandran}}\email{ravi@caltech.edu}

\affil*[1]{\orgdiv{Division of Engineering and Applied Science}, \orgname{California Institute of Technology}, \orgaddress{\city{Pasadena}, \postcode{91125}, \state{CA}, \country{USA}}}

\affil[2]{\orgdiv{Department of Engineering}, \orgname{University of Cambridge}, \orgaddress{\city{Cambridge}, \postcode{CB2 1PZ}, \country{UK}}}



\abstract{\textbf{Background: } Full-field, quantitative visualization techniques, such as digital image correlation (DIC), have unlocked vast opportunities for experimental mechanics. However, DIC has traditionally been a surface measurement technique, and has not been extended to perform measurements on the interior of specimens for dynamic, full-scale laboratory experiments. This limitation restricts the scope of physics which can be investigated through DIC measurements, especially in the context of heterogeneous materials. 
\unboldmath

\textbf{Objective:} The focus of this study is to develop a method for performing internal DIC measurements in dynamic experiments. The aim is to demonstrate its feasibility and accuracy across a range of stresses (up to $650\,$MPa), strain rates  ($10^{3}$-$10^6\,$s$^{-1}$), and high-strain rate loading conditions (e.g., ramped and shock wave loading).

\textbf{Methods:} Internal DIC is developed based on the concept of applying a speckle pattern at an inner-plane of a transparent specimen. The high-speed imaging configuration is coupled to the traditional dynamic experimental setups, and is focused on the internal speckle pattern. During the experiment, while the sample deforms dynamically, in-plane, two-dimensional deformations are measured via correlation of the internal speckle pattern. In this study, the viability and accuracy of the internal DIC technique is demonstrated for split-Hopkinson (Kolsky) pressure bar (SHPB) and plate impact experiments.

\textbf{Results:} The internal DIC experimental technique is successfully demonstrated in both the SHPB and plate impact experiments. In the SHPB setting, the accuracy of the technique is excellent throughout the deformation regime, with measurement noise of approximately $0.2\%$ strain. In the case of plate impact experiments, the technique performs well, with error and measurement noise of $1\%$ strain.

\textbf{Conclusion:} The internal DIC technique has been developed and demonstrated to work well for full-scale dynamic high-strain rate and shock laboratory experiments, and the accuracy is quantified. The technique can aid in investigating the physics and mechanics of the dynamic behavior of materials, including local deformation fields around dynamically loaded material heterogeneities.}




\maketitle

\section{Introduction}


Traditional non-contacting, qualitative imaging techniques, such as conventional imaging, shadowgraphy, Schlieren \cite{Settles2001schlieren}, and others have played a significant role in understanding the occurrence of many general mechanics phenomena, including characterizing deformation and failure mechanisms. However, these methods are inadequate to quantitatively capture local details of deformations. Full-field, quantitative imaging techniques have revolutionized the fields of experimental mechanics and mechanical behavior of materials, providing detailed insight into the stress or strain fields, especially in those experiments with complicated, non-uniform deformations. Some popular techniques have included photoelasticity \cite{Dally1980PhotoElasticity}, coherent gradient sensing (CGS) \cite{Tippur1991CGS}, and digital image correlation (DIC) \cite{Peters1982DIC,Sutton1983DIC,Schreier2009}, which provide powerful complements to traditional pointwise interferometric and other measurement techniques. DIC, which is currently the most popular choice among the mechanics community, is a methodology used to extract full-field displacement measurements based on pattern matching between grayscale images of the deformed and undeformed state of an object with a speckled surface \cite{Schreier2009}. While traditionally focused on quasi-static strain rates, recent advances in high-speed camera technology have enabled accurate implementation of DIC during high-strain rate dynamic experiments \cite{Bodelot2015Copper,Malchow2018Magnesium,Reu2008HighSpeedDIC,Seidt2017Synchronous,Keyhani2019novel}, and most recently has been extended to the shock compression regime via full-field free surface velocity measurements \cite{Ravindran2023ThreeDim}. True to its intended purpose of measuring spatially varying deformations, the DIC technique for shock compression experiments has even been extended to study the non-uniform shock structuring arising in plate impact experiments on particulate composite materials \cite{Ravindran2023Mesoscale}. Yet, it is recognized that there remains a need in the study of heterogeneous materials for \textit{in-situ} characterization of the dynamic deformation field at the source of the heterogeneity---i.e., at the material boundary/interface and at the length scale of the heterogeneity itself. These boundaries/interfaces have often been found to be the location of failure via delamination \cite{Wisnom2012Delamination}, fracture \cite{Abrate1998Composite}, dynamic instabilities (e.g., jetting \cite{Escauriza2020Collapse,Branch2017ControllingShock}), shear localization \cite{Lovinger2024Localization}, etc. Hence, accurate characterization of the deformation evolution at these boundaries is critical. Since these boundaries, such as those associated with defects (e.g., voids/pores and inclusions), are predominantly located inside of the matrix material, a DIC technique which enables these difficult measurements to be performed under dynamic loading conditions is much in need. Furthermore, for the purpose of capturing features at the small length scale of these heterogeneities, high-magnification imaging for this technique is often desired. Of particular interest are the problems of pore collapse and interaction between particles/fibers and matrix in composites. Besides use on heterogeneous materials, this technique would also be relevant for unraveling the three dimensional nature of experiments. Such scenarios, in which internal deformation measurements would be especially useful, include: (i) the state of deformation is neither plane strain nor plane stress, (ii) the physics of interest must be studied under confinement (e.g., lateral confinement of plate impact experiments), and (iii) the mechanics problem is sensitive to boundary effects. One interesting example is the study of internal crack propagation and interaction with the free surface.

 While DIC is traditionally a surface measurement, the internal DIC concept has been previously applied in a few investigations. Berfield, et al. were the first to demonstrate internal DIC in the quasi-static regime, and did so across the nm to $\mu$m length scales \cite{Berfield2007Micro}. They developed a methodology for internal DIC in polymers, demonstrated its accuracy, and applied it to investigate the deformation fields around silica micro-spheres embedded inside an elastomer. In addition, internal DIC has also been implemented in the dynamic strain rate regime for different loading conditions. For example, Huang, et al. \cite{Huang2015FailureWave} conducted plate impact experiments in conjuction with internal DIC to study properties of a failure wave in PMMA, though the accuracy of the technique was not verified. Recently, internal DIC has also been developed for dynamic visualization of laser-induced cavitation experiments in gels \cite{McGhee2023Microcavitation}, for which the specimen is dynamically loaded while situated under a microscope.

 The previous dynamic internal DIC endeavors have been restricted by the absence of comparison to ground truth data for the purpose of evaluating the techniques. Additionally, the experimental platforms have limited the physics which are accessible for investigation. This work addresses these issues by developing an internal DIC framework for use in traditional dynamic experiments such as split-Hopkinson (Kolsky) pressure bar (SHPB) and plate impact experiments, which possess well-prescribed loading conditions (uniaxial stress and uniaxial strain, respectively). This provides a suitable deformation (strain) state for comparison, against which the error is quantified. Additionally, the experimental setups used are general and adaptable to many loading conditions depending on the problem being addressed. Further, the internal DIC technique is not limited to these two experimental setups, but can be implemented in many full-scale laboratory experiments and should be applicable for optically transparent materials of interest.
 
 The experimental setup is discussed in \cref{sec:Methods}, with application to SHPB and plate impact experiments. In \cref{sec:Results}, validation experiments for both experimental setups are presented, the accuracy of the approach and corresponding experimental noise are quantified, and possible sources of error are identified. Finally, concluding remarks and future directions are given in \cref{sec:Conclusion}.

\section{Materials and Methods} \label{sec:Methods}
    The premise of the proposed internal DIC technique is to manufacture transparent target specimens with an internally embedded speckle pattern. Next, a series of deformation images are captured during dynamic compression of these target specimens using a high-speed imaging setup configured to visualize the internal speckle pattern through the transparent target. These images capture the in-plane displacements, which are computed through DIC, and strains, which are subsequently calculated from the displacement measurements. In this study, a series of split-Hopkinson (Kolsky) pressure bar (SHPB) \cite{Chen2010split} and plate impact \cite{Ramesh2008Experiments} experiments were conducted to demonstrate the internal DIC framework, and comparisons were made with experimental and theoretical measures to quantify the accuracy of the method. 
    \subsection{Sample Preparation}
\begin{figure*}[htpb]
            \centering
            \includegraphics[width=1.0\textwidth]{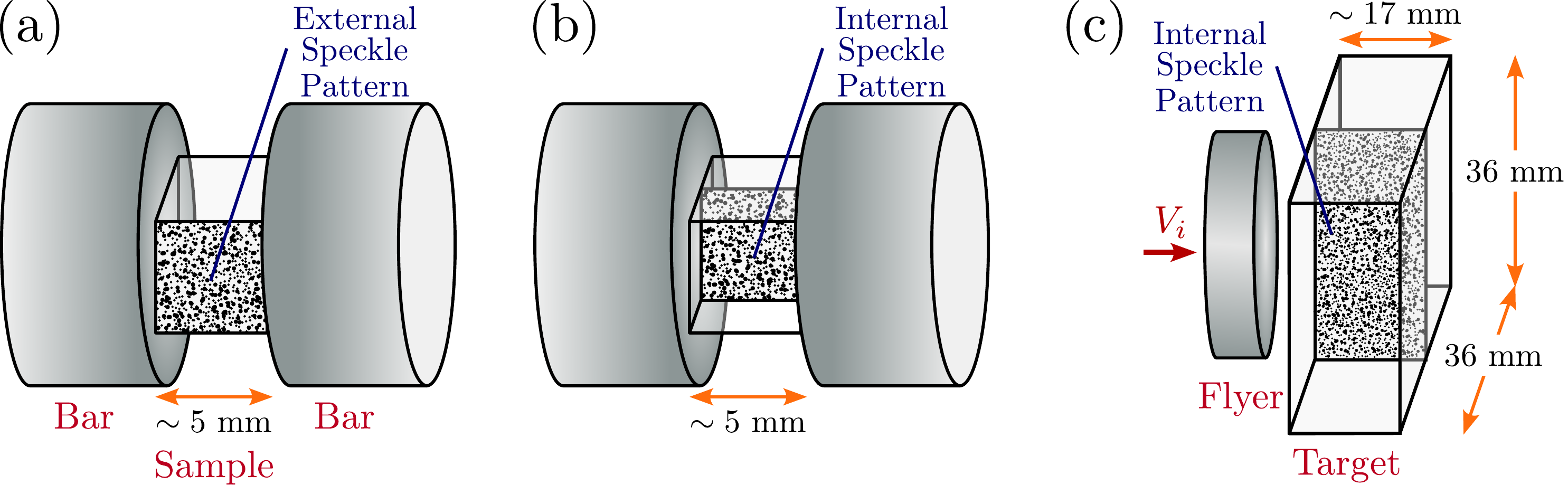}
            \caption{Specimen geometries: (a) Bulk cubic SHPB specimen which provides ground truth comparison. (b) Inner-plane cubic SHPB specimen used to validate the internal DIC technique. (c) Plate impact inner-plane specimen used to capture the shock response in PMMA and assess the accuracy of the internal DIC technique.}
            \label{fig:specimengeometries}
        \end{figure*}

        For the experiments presented, three different target specimen configurations were utilized, which are depicted in  \cref{fig:specimengeometries}. Each sample was manufactured from stock PMMA material obtained from E\&T Plastics (Long Island City, New York). \Cref{fig:specimengeometries}\hyperref[fig:specimengeometries]{a} shows a classical cubic sample used for SHPB experiments, for which the speckle pattern is applied to the external surface. This sample type is denoted as ``bulk''. Samples in \cref{fig:specimengeometries}\hyperref[fig:specimengeometries]{b-c} are used for internal DIC, and are manufactured out of two separate half-samples which are speckled and glued together, and will be called ``inner-plane'' samples. 

        The half-samples for inner-plane specimens shown in \cref{fig:specimengeometries}\hyperref[fig:specimengeometries]{b-c} are carefully lapped together in pairs to ensure identical dimensions, which aids precise gluing. Lapping also creates flat surfaces at the glue interface and load/impact surface, which are crucial to ensure a strong glue bond, uniform loading (for SHPB), and planar shock structure (for plate impact) during the experiments. Additionally, the half-sample surfaces are polished along the visualization direction to create maximum transparency for imaging. A speckle pattern, discussed in \cref{sec:DIC}, is then applied to the inner surface of one half-sample and allowed to dry for 24 hours, after-which the two pieces are sandwiched together with a thin layer of two-part epoxy glue, EpoxAcast 690 from SmoothOn (Macungie, Pennsylvania), at the speckle interface. The samples are precisely aligned using a confining apparatus, and weighed down to squeeze out excess glue. After the glue has set for the manufacturer-specified 24 hours, the weights are removed and the sample is taken out of the confining apparatus. Excess glue is removed from the sides of the sample, and the intact sample surfaces are re-lapped in the load/impact direction until the variation in thickness measurements is below $50 \, \mu$m for SHPB specimens, which ensures uniform load in the sample and improves repeatability of experiments. For plate impact specimens, the restriction is tightened to require less than $20 \, \mu$m thickness variation, and the surface flatness, as measured by Fizeau rings under monochromatic light, is required to be less than $1 \, \mu$m. These specifications ensure that a planar shock wave is generated upon impact during the plate impact experiments. Final assembled SHPB samples are approximately $5\times5\times5\,$mm cubes, while plate impact samples are $36\times36\,$mm square plates with approximately $17\,$ mm thickness.

        For plate impact experiments, circular flyer plates made of aluminum 7075 are also prepared, with typical dimensions of $35\,$mm diameter and $13\,$mm thickness. They are similarly lapped until they satisfy requirements of less than $10\,\mu$m thickness variation and less than $0.5\,\mu$m surface flatness on the impact side of the flyer.

        Electrical shorting pins are utilized in plate impact experiments to trigger diagnostics upon impact and simultaneously measure impact tilt (planarity). After the target has been prepared as described above, tilt pins are glued into four holes which correspond with the perimeter of the flyer. After gluing, these pins are sanded down and the surface is lapped one final time to ensure the pins are flush with the impact surface. Next, the pins are wired into a digital logic circuit whose output is connected to a high-speed digital oscilloscope, and the target is mounted onto the target holder, which is affixed to a six-degree-of-freedom gimbal. Lastly, the assembly is transferred to the vacuum chamber to begin the alignment process.

        \subsection{Split-Hopkinson Pressure Bar Experimental Setup}
\begin{figure}[h]
                \centering
                \includegraphics[width=0.8\linewidth]{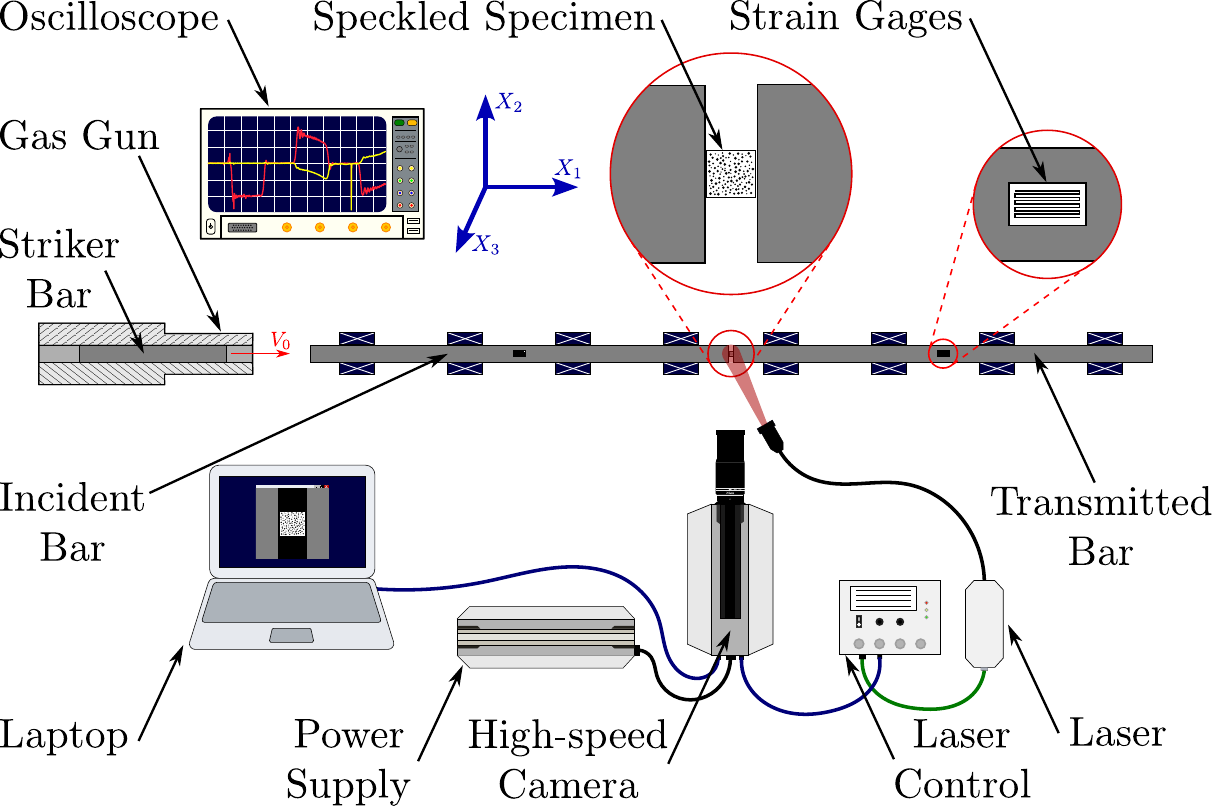}
                \caption{Experimental setup for split-Hopkinson (Kolsky) pressure bar (SHPB) experiments. The setup consists of the traditional components: gas gun, striker bar, incident bar, transmitted bar, strain gage system, oscilloscope, and sample. Additionally configured is the high-speed imaging setup which incorporates the high-speed camera and light source for imaging, in addition to speckle patterns used for DIC.}
                \label{fig:SHPBSetup}
            \end{figure}

        High-strain rate experiments were conducted using the split-Hopkinson pressure bar (SHPB) experimental apparatus \cite{Kolsky1949Investigation}. The technique relies upon the propulsion of a striker bar using a pressurized gas gun, which impacts an incident bar and generates a stress wave. The wave propagates through the incident bar into the sample, which is sandwiched between the incident and transmitted bars, and reverberates inside the sample. When using the appropriate bar material and sample geometry, dynamic equilibrium is quickly achieved. This enables traditional analysis of the sample's mechanical state via strain gage measurements on the bars, leveraging elastic wave theory and the assumption of force balance, i.e., specimen equilibrium \cite{Ramesh2008Experiments}. For this work, the conventional SHPB experimental apparatus is complemented by a high-speed imaging setup, which is configured to capture deformation images of the sample, which is depicted in \cref{fig:SHPBSetup}. Post-processing of the deformation images via DIC enables extraction of the strain evolution in the sample. The strain gage measurement system and high-speed imaging are simultaneously triggered when the stress wave is detected by the strain gages on the incident bar.
        
        In this work, aluminum 7075 bars are used, with the striker bar measuring $0.46\,$m in length, incident and transmitted bars both having a length of $1.83\,$m; all with diameter $19.05 \,$mm. Two strain gages, Omega SGD-2D/350-LY11 (Norwalk, Connecticut), are mounted at the midpoint of each bar (diametrically opposed to one another, to average out any bending strains), the voltage signal is conditioned by a Vishay 2310B signal conditioning amplifier (Malvern, Pennsylvania), and the signal is recorded with a four-channel, 1 GHz, Agilent MSO-X 4104A digital oscilloscope (Santa Clara, California). The high-speed imaging setup is composed of a Shimadzu HPV-X2 camera (Kyoto, Japan) equipped with an 100 mm Tokina AT-X Pro lens (Tokyo, Japan) and a Cavitar Cavilux incoherent laser illumination source (Tampere, Finland), which is setup to visualize the sample during dynamic compression. Details of the DIC speckle pattern and analysis are provided in \cref{sec:DIC}.
        
    \subsection{Plate Impact Experimental Setup}
   
        High-strain rate, normal plate impact experiments were conducted with a powder gun facility at Caltech, equipped with a $3\,$m long keyed barrel with an inner diameter of $38.7\,$mm. After sample preparation is finished and the target is mounted onto the six-degree-of-freedom gimbal, it is placed into the vacuum chamber for alignment. The flyer plate is glued to the projectile, which is placed into the end of the gun barrel. Next, the target plate is aligned in translational and rotational directions to minimize impact tilt. Then, the imaging diagnostics are set up, and the chamber is closed before final preparations are made to fire the projectile assembly.

\begin{figure*}[htpb]
            \centering
            \includegraphics[width=0.5\textwidth]{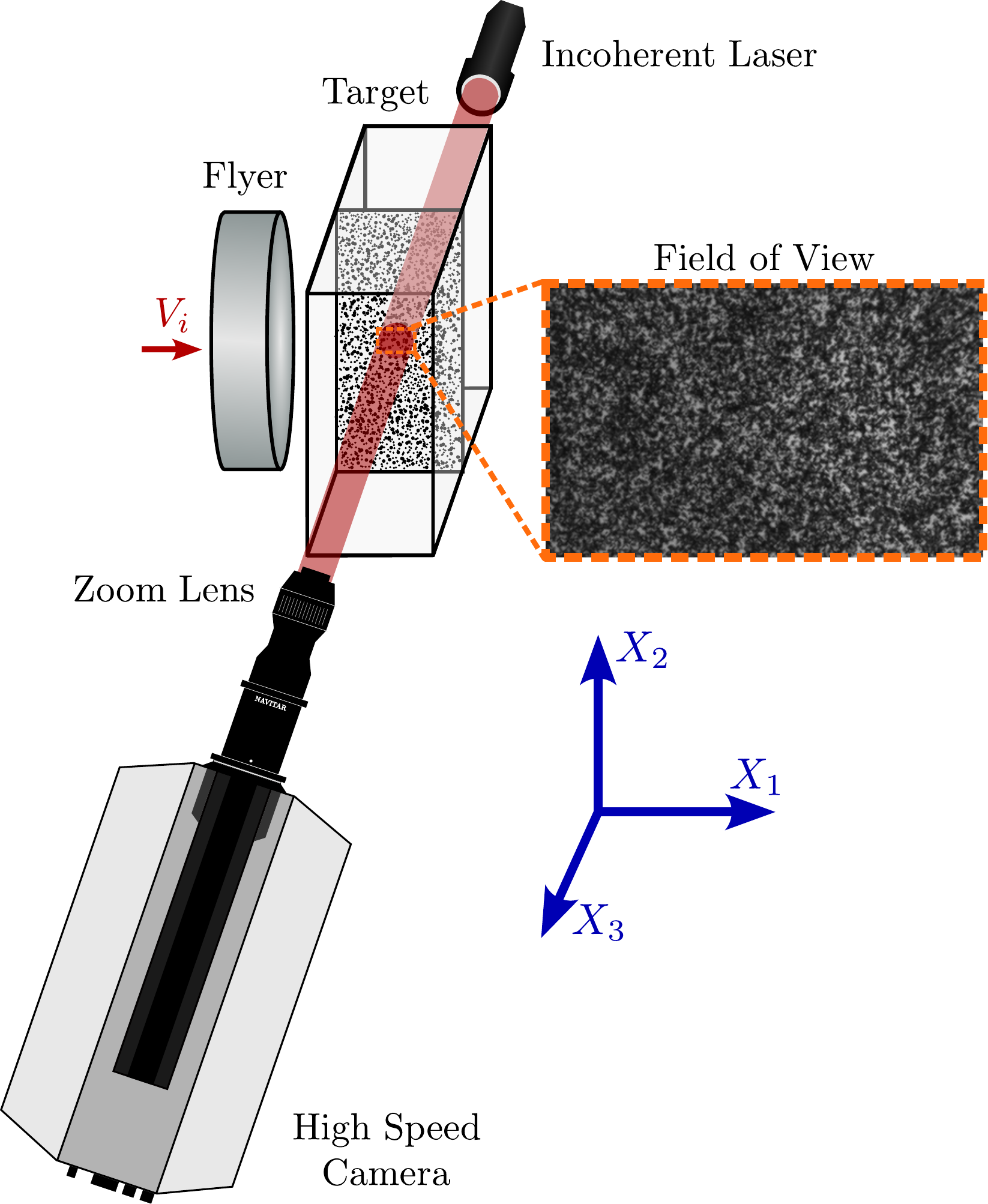}
            \caption{Experimental setup for plate impact experiments. Depicted is the flyer plate approaching the target. Also shown is the high-speed, high-magnification imaging setup, complete with backlit laser illumination, and an example field of view (see inset).}
            \label{fig:PlateImpactSetup}
        \end{figure*}

        The impact event is generated by igniting a gun-powder charge immediately behind the projectile. The resulting pressure build-up accelerates the projectile down the barrel, until it makes planar contact with the target specimen in the vacuum chamber. Upon impact, tilt pins in the target are shorted, triggering the diagnostics and measuring the impact time at four locations on the target. Based on time of impact and measured impact velocity, the angle between the flyer and target at impact (tilt) is calculated. Simultaneously, a shock wave initiates at the impact interface in both the flyer and the target, the effect of which is captured by the diagnostics. Classical experiments utilize free surface velocity measurements at the rear of the sample, along with one-dimensional shock wave theories, to extract the material response under shock compression. Instead, here, high-speed imaging is used to visualize the shock propagation from left to right in the field of view, along with the resulting deformation behind the shock. The details of this visualization approach are discussed in \cref{sec:DIC}, and a schematic of the experimental setup is shown in \cref{fig:PlateImpactSetup}. In addition to high-speed imaging, which captures the deformation in the target, the impact velocity is measured via two precisely spaced laser gates which measure the difference in time at which the projectile passes each laser and breaks the laser gates, just prior to impact.

    \subsection{High-speed imaging and Digital Image Correlation (DIC)} \label{sec:DIC}
        
        While the high-speed imaging setup in the SHPB configuration follows conventional practices, translation of such a setup to plate impact experiments introduces several challenges. The same camera (Shimadzu HPV-X2) and light source (Cavitar incoherent laser) are used, but a number of issues must be addressed, including higher framing rates, in-material refractive index changes under shock compression, and the addition of high-magnification imaging. These details are discussed below, in addition to the DIC speckling and post-processing.

        To begin with, high-magnification imaging was performed for the plate impact validation to ensure the internal DIC technique is applicable at fine length scales. This is relevant for many heterogeneous materials (e.g., porous media with pores on the order of $\mu$m$-$mm), and also provides sufficient DIC resolution to capture strain localization which may occur during dynamic loading. This was accomplished by using a Navitar $0.7-4.5$x zoom lens and 2x adapter tube assembly (Rochester, New York), which achieved a typical field of view of $2.8\times1.75\,$mm ($400\times250\,$pixels, $7\,\mu$m/pixel). Compared to the SHPB experiments, for which the field of view in each experiment was approximately $7.2\times4.5\,$mm ($18\,\mu$m/pixel), this is $2.6$ times higher magnification. However, this gives rise to several issues. First, because of significant lens curvature inherent in the zoom lens, distortions are introduced to the images. This is remedied by taking a series of images in which the specimen undergoes rigid body motion in both horizontal and vertical translation directions, and applying a distortion correction function to regain the uniform displacement field. The same correction function can then be applied to each deformation image during the experiment. Second, the framing rate is increased from $2,000,000\,$fps (SHPB) to $10,000,000\,$fps (plate impact). This ultra-high speed imaging, in conjunction with the high magnification, creates a severely light-starved situation. Typically, the synchronized pulsed laser illumination source ($50\,$ns pulse) used in this work provides sufficient light at this framing rate \cite{Ravindran2023ThreeDim,Ravindran2023Mesoscale}, but here it's additionally necessary to configure the light source in a back-lit configuration (\cref{fig:PlateImpactSetup}) to maximize light captured by the zoom lens.

        Preliminary experiments also revealed that when the shock wave passes through the field of view, the steep density gradient causes temporary loss of transparency in the PMMA, which is quickly recovered behind the shock wave. However, an artificial rigid body motion was observed behind the wave, owing to the sharp change in refractive index of the material across the shock wave. Because the visualization plane is inside the PMMA target which is undergoing shock compression (i.e., light must carry information through the PMMA before arriving at the camera), this change in refractive index can cause optical distortions. The distortion can be mitigated by aligning the camera such that the lens is parallel to the target window surface (the outer, polished surface of the target which is parallel to the speckle plane). To this end, mirrors are mounted to the target window surface and extension tube, alignment is checked using an auto-collimator, and the alignment is fine-tuned using a five-degree-of-freedom optical stage for the camera. In theory, when these are perfectly aligned, the refractive index change would have no influence on the deformation images; in practice, there remains a small distortion which is primarily uniform (hence minimal influence in the strain measurements). The error associated with the technique is quantified through validation experiments in \cref{sec:Results}.

        Much attention, rightly, goes to the careful setup of the visualization system. However, the speckle pattern and the post-processing procedures also play a significant role in quantitative visualization. For each SHPB experiment (except IP4, \cref{tab:Hopkinson Shot Summary}), the speckle pattern is applied with an airbrush, applying black speckles onto a thin, white, spray-painted background layer. The airbrush patterning generated suitable speckle sizes for low magnification imaging ($30-100\,\mu$m/speckle and $18\,\mu$m/pixel). In preparation for the plate impact experiment, toner powder was used in place of airbrushed black paint for one SHPB experiment (IP4) to ensure that black paint and toner powder are interchangeably non-intrusive for inner-plane specimens. For plate impact experiments, the toner powder was suspended inside a transparent paint which was airbrushed onto the internal surface. This method generated $10-20\,\mu$m speckles, which is ideal for high-magnification DIC with a resolution of $7\,\mu$m/pixel.

        After capturing experimental deformation images, DIC post-processing \cite{Schreier2009} was carried out using Correlated Solutions Vic-2D software (Columbia, South Carolina). Prior to analysis, distortion correction was carried out with the built-in correction algorithm. DIC analysis extracted in-plane (two-dimensional) full-field displacements from the deformation images using a subset size of 21 pixels along with a step size of 1 pixel. These DIC correlation settings were used for all experiments presented in this study. Following this, full-field strains were computed through discrete differentiation of the displacement field. For all SHPB experiments presented, the strain is computed in the Vic-2D software as the Lagrangian strain, \begin{equation} \varepsilon_{ij} = \frac{1}{2}\left(\frac{\partial u_i}{\partial X_j} + \frac{\partial u_j}{\partial X_i} + \frac{\partial u_k}{\partial X_i}\frac{\partial u_k}{\partial X_j}  \right)  \label{eq:Lagrangian Strain} \end{equation} where $u_i$ is the displacement component in the $X_i$ direction and repeated indices indicate summation. Additionally, an inherent 90\% center-weighted Gaussian spatial filter with a 15 pixel filter size is applied. Alternatively, for the plate impact experiment, to present the closest comparison to one dimensional shock theory, the one dimensional engineering strain (ignoring higher order terms in the Lagrangian strain metric) is computed below with a uniform spatial filter with a 15 pixel filter size. \begin{equation} \varepsilon_{11}^{Eng.} = \frac{\partial u_1}{\partial X_1} \label{eq:Engineering Strain} \end{equation}

\section{Results} \label{sec:Results}
    Validation experiments for the internal DIC technique were performed, first for SHPB experiments at two different strain rates, and then for plate impact experiments at a selected impact stress. The purpose of these experiments is three-fold: (i) to ensure the inner-plane specimen geometry does not introduce a non-physical material response, (ii) to confirm the ability to capture internal deformation fields using the proposed technique, and (iii) to quantify errors associated with the technique.
    
    \subsection{Split-Hopkinson Pressure Bar Experiments}
    
        SHPB experiments were conducted at strain rates of approximately $1500$ and $3700\,$s$^{-1}$, corresponding to impact velocities of approximately $11.6$ and $19.6\,$m/s, respectively, with two bulk SHPB (\cref{fig:specimengeometries}\hyperref[fig:specimengeometries]{a}) and two inner-plane SHPB (\cref{fig:specimengeometries}\hyperref[fig:specimengeometries]{b}) samples tested at each strain rate (impact velocity). In these experiments, dynamic equilibrium is achieved quickly, generating a nominally uniaxial stress loading condition. Hence, the deformation response should be uniform through the thickness of the sample, and the internal DIC measurements from inner-plane samples can be directly compared to the surface DIC measurements from bulk samples under nearly identical loading conditions (same impact velocity). The high-speed imaging was conducted at $2\,$million fps with a resolution of $18\,\mu$m/pixel to attain a field of view of $7.2\times4.5\,$mm ($400\times250\,$pixels).

    \begin{table*}[htpb]
        \setlength{\tabcolsep}{7.5pt}
        \centering
        \caption{\label{tab:Hopkinson Shot Summary} Summary of SHPB experiments.}
        \makebox[\textwidth][c]{
        \resizebox{1.3\textwidth}{!}{
        \begin{tabular}{c c c c c c c} \hline\hline
        Experiment & Sample & Impact & Speckle & \multicolumn{3}{c}{\textbf{Sample Dimensions}*}\\
        Number & Type & Velocity (m/s) & Type & \multicolumn{1}{c}{L$_{1}$ (mm)} & \multicolumn{1}{c}{L$_{2}$ (mm)} & \multicolumn{1}{c}{L$_{3}$ (mm)} \\
        \hline
        B1 & Bulk & $11.7$ & Airbrush & $4.529 \pm 0.003$ & $5.001 \pm 0.040$ & $4.805 \pm 0.004$\\
        B2 & Bulk & $11.6$ & Airbrush & $4.548 \pm 0.003$ & $4.837 \pm 0.003$ & $4.838 \pm 0.003$\\ 
        B3 & Bulk & $19.7$ & Airbrush & $4.532 \pm 0.003$ & $4.812 \pm 0.005$ & $4.557 \pm 0.005$\\
        B4 & Bulk & $19.7$ & Airbrush & $4.564 \pm 0.002$ & $4.548 \pm 0.005$ & $5.021 \pm 0.031$\\
        IP1 & Inner-Plane & $11.7$ & Airbrush & $4.814 \pm 0.003$ & $4.845 \pm 0.006$ & $5.003 \pm 0.004$\\
        IP2 & Inner-Plane & $11.6$ & Airbrush & $4.772 \pm 0.005$ & $4.801 \pm 0.024$ & $4.990 \pm 0.005$\\
        IP3 & Inner-Plane & $19.7$ & Airbrush & $4.648 \pm 0.004$ & $4.539 \pm 0.019$ & $4.911 \pm 0.007$\\
        IP4 & Inner-Plane & $19.7$ & Toner Powder & $4.780 \pm 0.004$ & $4.652 \pm 0.016$ & $4.965 \pm \text{NA}$\\
        \hline\hline
        \end{tabular}}}
        \begin{tablenotes}[flushleft]\footnotesize
            \item *L$_1$ is the thickness of the sample in the direction of compression. L$_2$ and L$_3$ are the dimensions of the cross-section, corresponding to the coordinate system in \cref{fig:SHPBSetup}.
        \end{tablenotes}
    \end{table*}    
    
\begin{figure*}[htpb]
            \centering
            \includegraphics[width=1\textwidth]{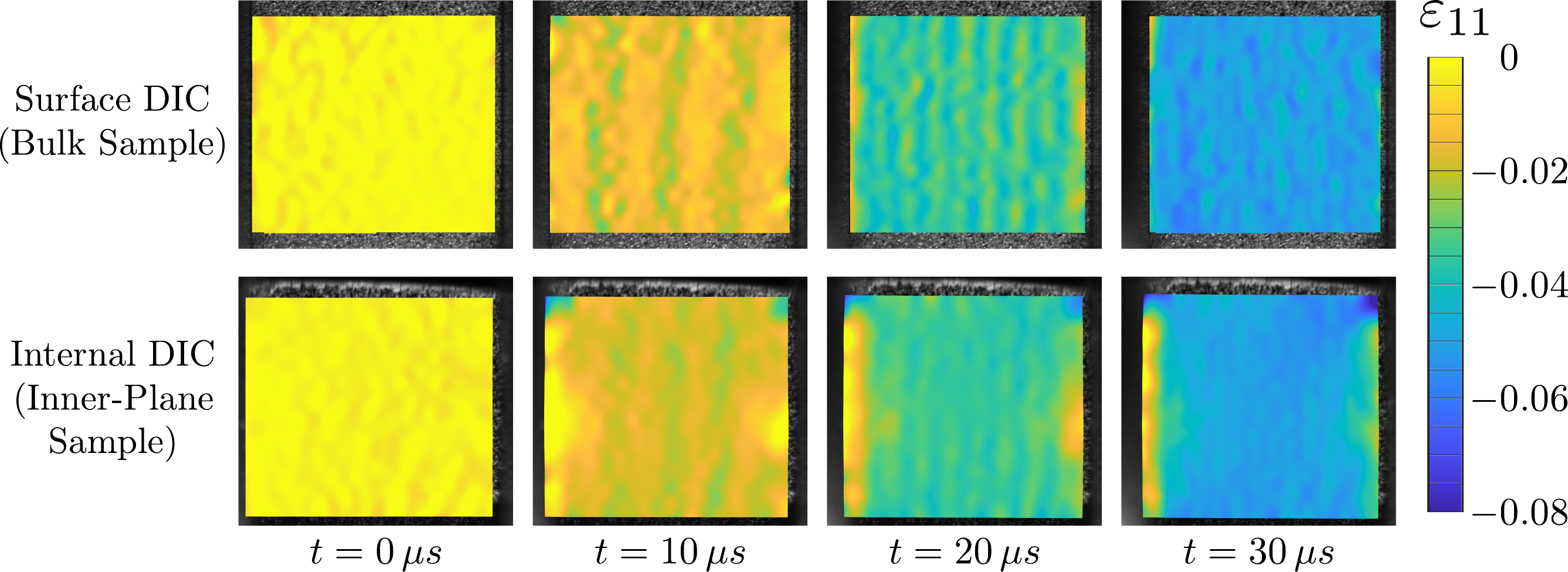}
            \caption{Frame-by-frame visualization of longitudinal strain on the  surface and inner-plane of PMMA samples under SHPB loading at identical striker bar impact velocity, 11.7 m/s. Experiments B1 and IP1 are shown on top and bottom, respectively. Time, t = 0 corresponds to the arrival of the loading pulse at the interface between the incident bar and the front of the specimen. The overlaid DIC strain fields are each approximately $4.5 \times 4\,$mm.}
            \label{fig:Bulk vs Inner Plane Hopkinson}
        \end{figure*}
\begin{figure*}[htpb]
            \centering
            \includegraphics[width=0.85\textwidth]{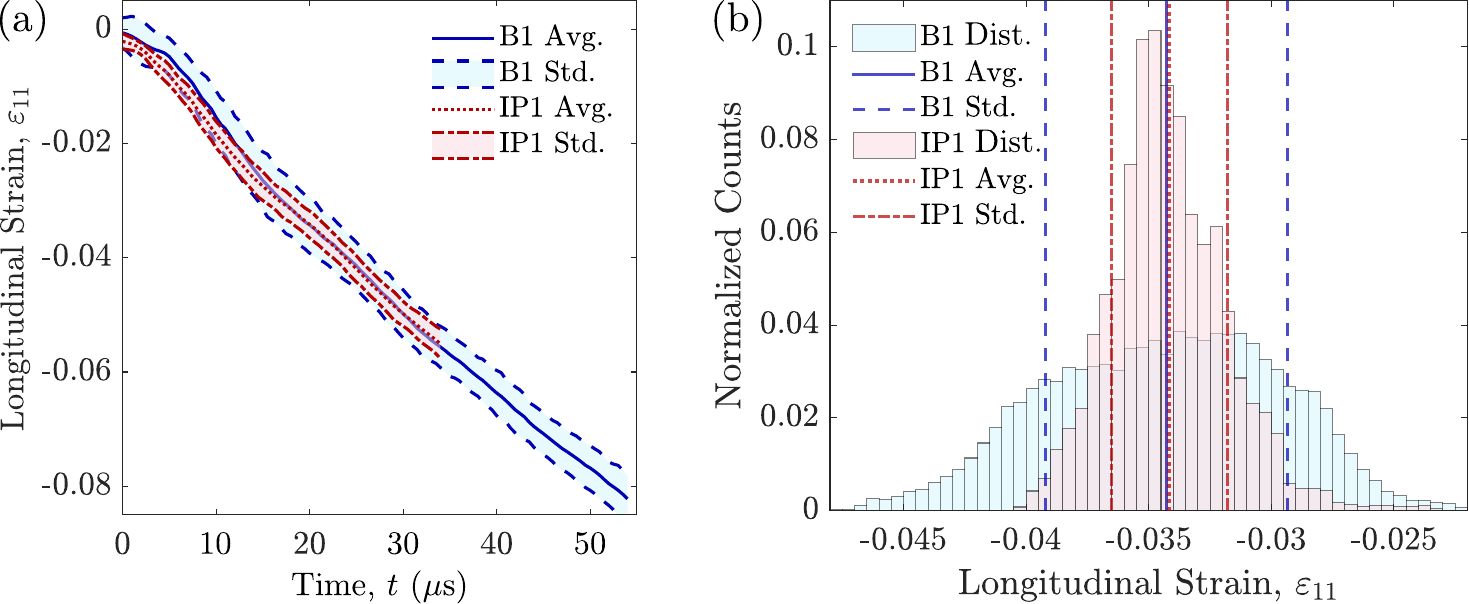}
            \caption{Error analysis of SHPB experiments B1 and IP1. Results for other experiments are similar and are omitted for the sake of clear visualization. (a) Longitudinal strain evolution, averaged over the area of interest, with shaded bounding curves representing one standard deviation from the mean. (b) Full histogram comparison at one time instance, $t=20\,\mu$s, comparing the spatial distribution of longitudinal strain measurements for B1 and IP1. Note the spread is largely dependent on the quality of the speckle pattern. In general the inner-plane samples possess equal or better error bounds when compared with bulk samples for the SHPB experiments.}
            \label{fig:SHPB Error}
        \end{figure*}

        In total, eight SHPB experiments are presented, the details of which are summarized in \cref{tab:Hopkinson Shot Summary}. An example comparison between experiments B1 (bulk) and IP1 (inner-plane) is shown in \cref{fig:Bulk vs Inner Plane Hopkinson} with a time series of selected images and overlaid DIC strain ($\varepsilon_{11}$) fields. Both the experiments were conducted under nearly identical conditions, with a striker bar impact velocity of $11.7\,$m/s. By visual inspection, both experiments show uniformity and minimal noise in the strain field. In addition, there is excellent agreement between the experiments, with the only notable difference being slight non-uniformity near the front and back faces of the sample, which are in contact with the incident and output bars, respectively. This feature is present in both B1 and IP1, but more significant in the inner-plane experiment (IP1). One can quantify the agreement, uniformity, and noise by investigating the strain evolution with time inside a $3\,$mm diameter circular region at the center of the sample, shown in \cref{fig:SHPB Error}\hyperref[fig:SHPB Error]{a}. The average longitudinal strains $(\varepsilon_{11})$ of the two experiments coincide very closely with one another, and the error bounds, denoting one standard deviation from the mean, are shown to be small. The distribution of full-field strain measurements for experiments B1 and IP1 is further visualized for one arbitrary time instance, $t=20 \, \mu$s, in \cref{fig:SHPB Error}\hyperref[fig:SHPB Error]{b}. In this particular comparison (experiments B1 and IP1), the inner plane experiment is observed to be less noisy than the bulk; however, this result proves to be largely dependent on the quality of the speckle pattern and lighting of the particular experiment. In general, when comparing all SHPB experiments, the inner-plane DIC measurements have approximately equal levels of noise to the bulk DIC measurements The results in experiments B1 and IP1 represent the largest noise among the eight experiments.
    
\begin{figure*}[htpb]
            \centering
           \makebox[\textwidth][c]{\includegraphics[width=1.3\textwidth]{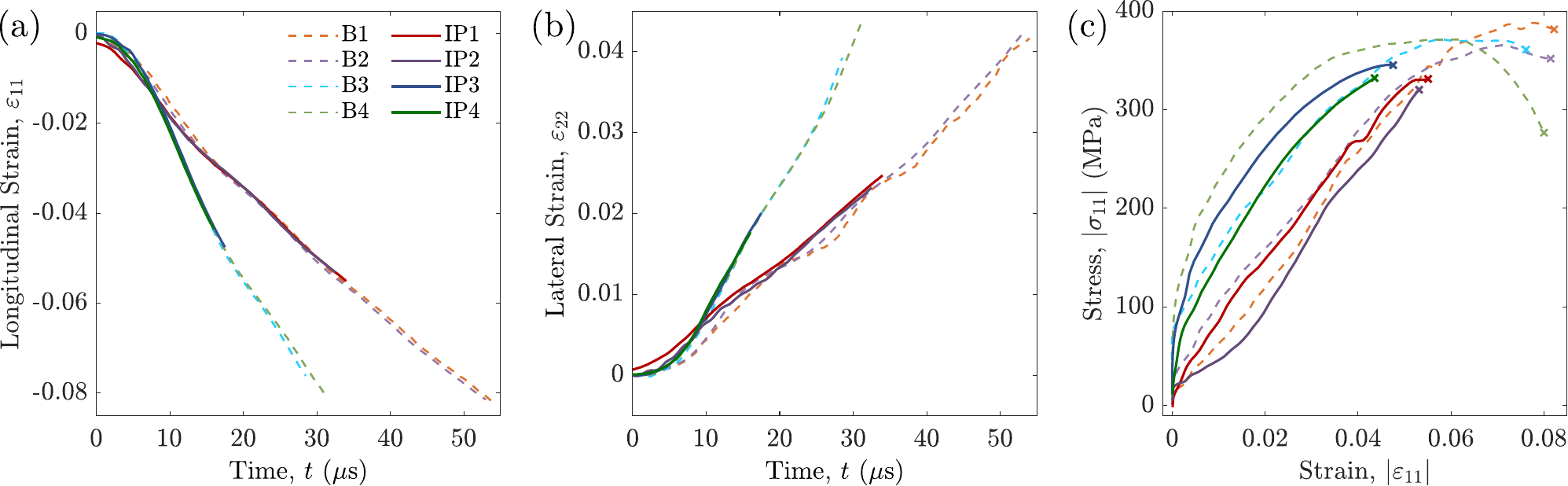}}
            \caption{Summary of all eight SHPB experiments. (a) Average longitudinal strain evolution, (b) Average lateral strain evolution, (c) Stress-strain curves. One can observe the close agreement of the strain evolution for each repeated impact velocity, regardless of specimen type (bulk or inner-plane). Stress-strain curves also show reasonable agreement during initial deformation (prior to failure). Note that the stress-strain curves are composed of DIC measurements of strain and strain gage measurements of stress.}
            \label{fig:Bulk vs Inner Plane Hopkinson Plots}
        \end{figure*}

        A similar quantitative comparison among all eight experiments is made in \cref{fig:Bulk vs Inner Plane Hopkinson Plots}, neglecting to include error bars for the sake of visibility, in which the time evolution of longitudinal $\left(\varepsilon_{11}\right)$ and lateral $\left(\varepsilon_{22}\right)$ strains, as well as the traditional stress-strain response, are compared. The averaged strain response from DIC is plotted in \cref{fig:Bulk vs Inner Plane Hopkinson Plots}\hyperref[fig:Bulk vs Inner Plane Hopkinson Plots]{a-b}, while the stress-strain response shown in \cref{fig:Bulk vs Inner Plane Hopkinson Plots}\hyperref[fig:Bulk vs Inner Plane Hopkinson Plots]{c} is derived through a combination of DIC strain measurements and engineering stress measurements via strain gage recordings on the transmitted bar, calculated as \begin{equation} \sigma(t) = \frac{A_{b}}{A_{s}}E_{b}\varepsilon_{T}(t).  \label{eq:SHPB Stress} \end{equation} Here, $A_{b}$ and $A_{s}$ are the cross-sectional areas of the bar and sample, respectively, $E_{b}$ is the Young's modulus of the bar material (aluminum 7075) ($E_{b} = 71.7\,$GPa), and $\varepsilon_{T}(t)$ is the strain measurement on the transmitted bar. One observes close agreement (among experiments with the same impact velocity) in the longitudinal and lateral strain measurements, indicating the technique works exceptionally well in the SHPB regime to characterize material deformation. However, at large strains the specimens begin to fail through brittle fracture. This is where notable differences arise, with the inner-plane specimens fracturing earlier than their bulk counterparts, because the internal DIC and glue interface supplies nucleation sites for fracture to occur. The stress-strain curves (\cref{fig:Bulk vs Inner Plane Hopkinson Plots}\hyperref[fig:Bulk vs Inner Plane Hopkinson Plots]{c}) make this difference clear, as the end of the curves, which mark the failure of the material, show consistent failure in the inner plane specimens at $4.4-5.5\%$ longitudinal strain, while the bulk specimens endure $7.6-8.2\%$ strain before failure. While this does pose a limitation on the technique for alternative applications, the current interest in these experiments lies only in the homogeneous deformation regime, not the failure regime.

        In the following section the technique is extended to plate impact experiments, which feature a uniaxial strain state of deformation and maintain an overall compressive loading state due to lateral confinement, thus preventing the occurrence of brittle fracture. Hence, the SHPB experiments provide sufficient proof of concept to transition the technique from SHPB to plate impact. For applications interested in the failure regime of the SHPB experiments, alternative sample preparation techniques should be implemented which incorporate an internal speckle pattern without introducing a material interface. One such technique is the embedded speckle plane patterning technique \cite{Berfield2007Micro,McGhee2023Microcavitation}, though it has limitations for the study of pre-existing heterogeneities.

    \subsection{Plate Impact Experiment}
    
        To validate the effectiveness of the internal DIC for shock experiments, a normal plate impact experiment is conducted at $0.64\,$GPa impact stress on a PMMA target plate, with an internal speckle pattern, by impacting the target with an aluminum 7075 flyer. The target's internal speckle pattern was imaged via high-speed imaging at a framing rate of 10 Mfps with an image resolution of $7\,\mu$m/pixel and field of view of $2.8\times1.75\,$mm ($400\times250\,$pixels). The experimental parameters for the experiment are summarized in \cref{tab:Plate Impact Shot Summary}.

        \begin{table*}[ht]
            \setlength{\tabcolsep}{7.5pt}
            \centering
            \caption{\label{tab:Plate Impact Shot Summary} Summary of normal plate impact experiment.}
            \makebox[\textwidth][c]{
            \resizebox{1.3\textwidth}{!}{
            \begin{tabular}{c c c c c c c c} \hline\hline
             Flyer          & Target        &       Flyer               & Target               & \phantom{*}L$_{\text{FOV}}$*    & Impact Velocity   & Impact Stress          & Tilt\\
             Material       & Material      &       Thickness (mm)      & Thickness (mm)       & (mm)                               & $V_0$ (m/s)       & $\sigma_{11}$ (GPa)   & (mrad)\\
            \hline
            Al 7075         & PMMA          & $12.799 \pm 0.001$        & $16.089 \pm 0.014$   & 5.0                                & $219 \pm 5$      & $0.64 \pm 0.02$       & 3.4 \\
            \hline\hline
            \end{tabular}}}
                \begin{tablenotes}[flushleft]\footnotesize
                \item *L$_{\text{FOV}}$ indicates the distance from the impact surface to the center of the field of view (FOV).
                \end{tablenotes}
        \end{table*}
        
        For plate impact experiments, assuming uniaxial strain deformation, one can compute a theoretical shock strain \cite{Meyers1994DBOM} against which the experimental result can be compared: \begin{equation} \varepsilon_{11}^{\text{Theory}} = \frac{u_p}{U_s}. \label{eq:ShockStrain} \end{equation}

        \noindent The particle velocity $\left(u_p\right)$ can be computed from the measured impact velocity $\left(V_0\right)$ and empirically known equations of state (EOS) relating the shock velocity $\left(U_s\right)$ to the particle velocity ($U_s-u_p$ relation). This well established methodology is known as impedance matching, which is well summarized by Meyers \cite{Meyers1994DBOM}. Tabular EOS data, for use in this technique, is readily available for both the flyer and target materials, aluminum 7075 \cite{Marsh1980LASL} and PMMA \cite{Barker1970PMMA}, respectively. The linear $U_s-u_p$ relation (\cref{eq:EOS}) has been fit to the tabular data for each material, and the fitted material parameters, $C_0$ and $S$, are shown in \cref{tab:Properties}.   \begin{equation} \label{eq:EOS} U_s = C_0 + Su_p \end{equation}

        \begin{table*}[ht]
 	\centering
 	\begin{threeparttable}
 		\setlength{\tabcolsep}{7.5pt}
 		\caption{Equation of state parameters.}
 		\centering
	\begin{tabular}{cccc}
		\hline \hline
		Material                                & \begin{tabular}{@{}c@{}} Density, $\rho_0$ \\ {[}kg/m$^3${]}\end{tabular}   &   \begin{tabular}{@{}c@{}}$C_0$ \\ {[}m/s{]}\end{tabular}                 & $S$ \\ \hline
		PMMA \cite{Barker1970PMMA}              & 1186                                                                        &   2770                                                                    & 2.11  \\
            Aluminum 7075 \cite{Marsh1980LASL}      & 2804                                                                        & 5022                                                                  & 1.99   \\ 
        \hline\hline                          
	\end{tabular}
	\label{tab:Properties}
 	\end{threeparttable}
        \end{table*}
    
\begin{figure*}[htpb]
            \makebox[\textwidth][c]{\includegraphics[width=1.3\textwidth]{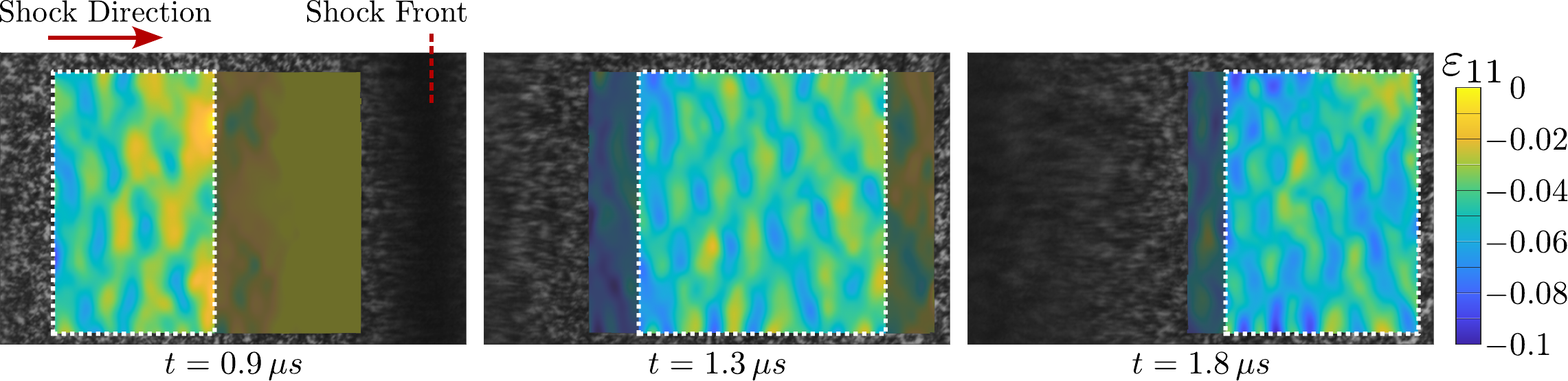}}
            \caption{A series of images from the plate impact experiment, with longitudinal engineering strain, $\varepsilon_{11}$, overlaid. Time is shifted such that $t=0$ coincides with the arrival of the shock wave in the field of view. The area of interest (region free from influence of wave distortion) is marked by the white, dashed box.}
            \label{fig:Validation PI Frames}
        \end{figure*}

        To evaluate the performance of the technique for plate impact experiments, one begins by considering the full-field longitudinal strain measurements. \Cref{fig:Validation PI Frames} shows selected time instances during the experiment, where the area of interest (AOI) is marked by a white dashed box. This area of interest is determined by carefully selecting the front and back boundaries through inspection of the raw deformation images: selecting only the regions which are free of blurriness, which distorts the DIC-computed strain fields. Blurriness arises, in part, from the steep density gradients associated with the shock wave (strong discontinuity in stress) and release waves which follow after the shock. In this experiment the shock wave, which travels from left to right, arrives in the field of view (FOV) at time, $t=0$, and can be seen exiting the FOV at $t=0.9\,\mu$s. Similarly, a blurry front enters the FOV at $t=1.3\,\mu$s and covers half the FOV at $t=1.8\,\mu$s. It is not clear whether this blurry front arises due to release waves from the impact surface or boundaries of the target, or from another source. Alternative sources could include lateral relaxation at the boundary of the target plate or changing refractive index due to changes in the stress state in the PMMA material which the camera looks through. Both of these sources would contribute to shifting the optical focal plane, which would introduce blur or defocusing. Motion blur due to large velocities is not an issue, which is evidenced by the crisp image quality immediately behind the shock wave where peak particle velocity is achieved.
        
\begin{figure}[h]
            \makebox[\textwidth][c]{\includegraphics[width=1.3\textwidth]{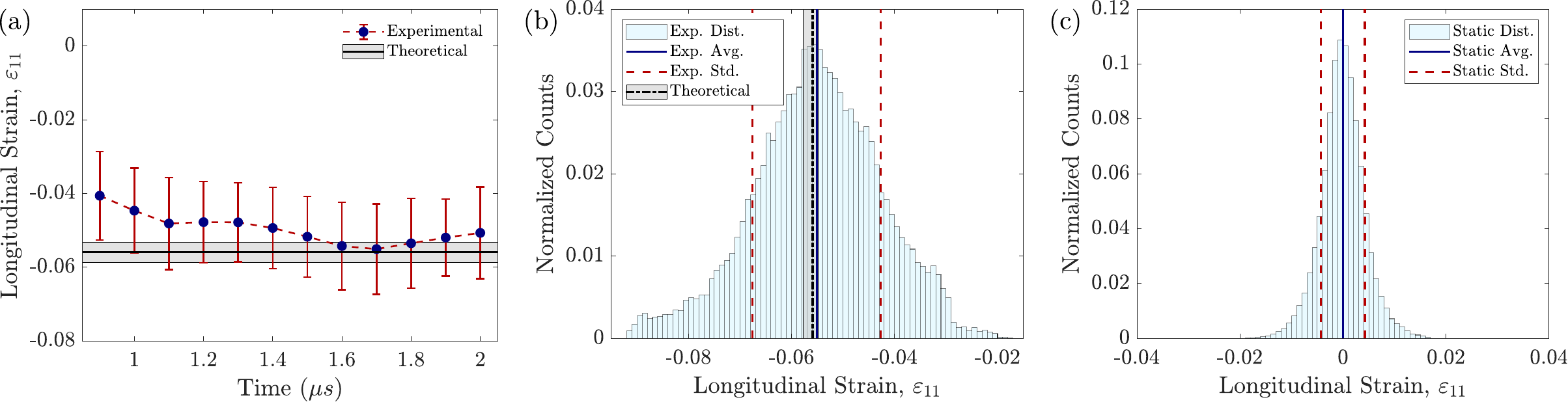}}
            \caption{DIC results for the plate impact experiment inside the area of interest. (a) Averaged experimental longitudinal strain response compared with the theoretical shock strain (\cref{eq:ShockStrain}) for the given impact conditions and material choice. Time is shifted such that $t=0$ coincides with the arrival of the shock wave in the field of view. Error bars represent one standard deviation from the mean. (b) Histogram of local longitudinal strain measurements for a given time instance, $t=1.7\,\mu$s, depicting the full measurement distribution in the area of interest. (c) Histogram of local longitudinal strain measurements for several static, rigid body translation images, using the same speckle pattern as in the experiment, to provide a noise floor for the DIC setup.}
            \label{fig:Validation PI Plot}
        \end{figure}

        Examining the computed longitudinal strain field, it is apparent that within the AOI at each time step, the strain fields are fairly uniform, albeit with relatively large noise. This uniformity is an important feature, as the Hugoniot steady shocked state should possess constant, uniform strain behind the shock wave. Further, the averaged strain response in the AOI is displayed in \cref{fig:Validation PI Plot}\hyperref[fig:Validation PI Plot]{a} where the steady shocked state is shown to be nearly constant with respect to time, and approaches the theoretical shock strain. For the majority of the time instances, the error is less than 1\% strain. 

        This experimental deviation from the theoretically predicted shock strain---measuring less than the theoretical value---is evidence of slight systemic error which may be attributed to small optical distortions arising from the deformation of the transparent PMMA target which is being imaged through. The error may also be caused by possible deviation from the idealized uniaxial strain loading conditions. Additional noise is also inherent in these results, caused by several sources including DIC artifacts, rigid body motion arising from the temporary loss of correlation at the shock wave location, high-magnification imaging distortions, other optical distortions, and pixel interpolation used by the Shimadzu HPV-X2 camera at 10 Mfps imaging rate. To decouple the effects of the experimental configuration (e.g., camera, speckle pattern, DIC processing scheme, etc.) from those of the effects of shock compression (e.g., optical distortions), the strain noise should thus be compared with that of static, rigid body translation. The standard deviation for longitudinal strain is $0.0043$ for static rigid body translation and $0.0116$ under shock, meaning the experimental noise under shock compression is $2.7$ times larger than that of ambient, static images. This result is visualized through a comparison of histograms for the data inside the AOI for a given time instance in the experiment (\cref{fig:Validation PI Plot}\hyperref[fig:Validation PI Plot]{b}) and for static, rigid body translation images (\cref{fig:Validation PI Plot}\hyperref[fig:Validation PI Plot]{c}). While not negligible, these error levels are deemed acceptable, especially for applications involving large, localized strains such as the shock compression of heterogeneous materials. The error can also be lowered through additional filtering (when suitable for the application) and through refinements of the technique such as improved speckle patterning, alignment, lighting, and camera hardware. 

        \begin{figure}[h]
            \makebox[\textwidth][c]{\includegraphics[width=1.0\textwidth]{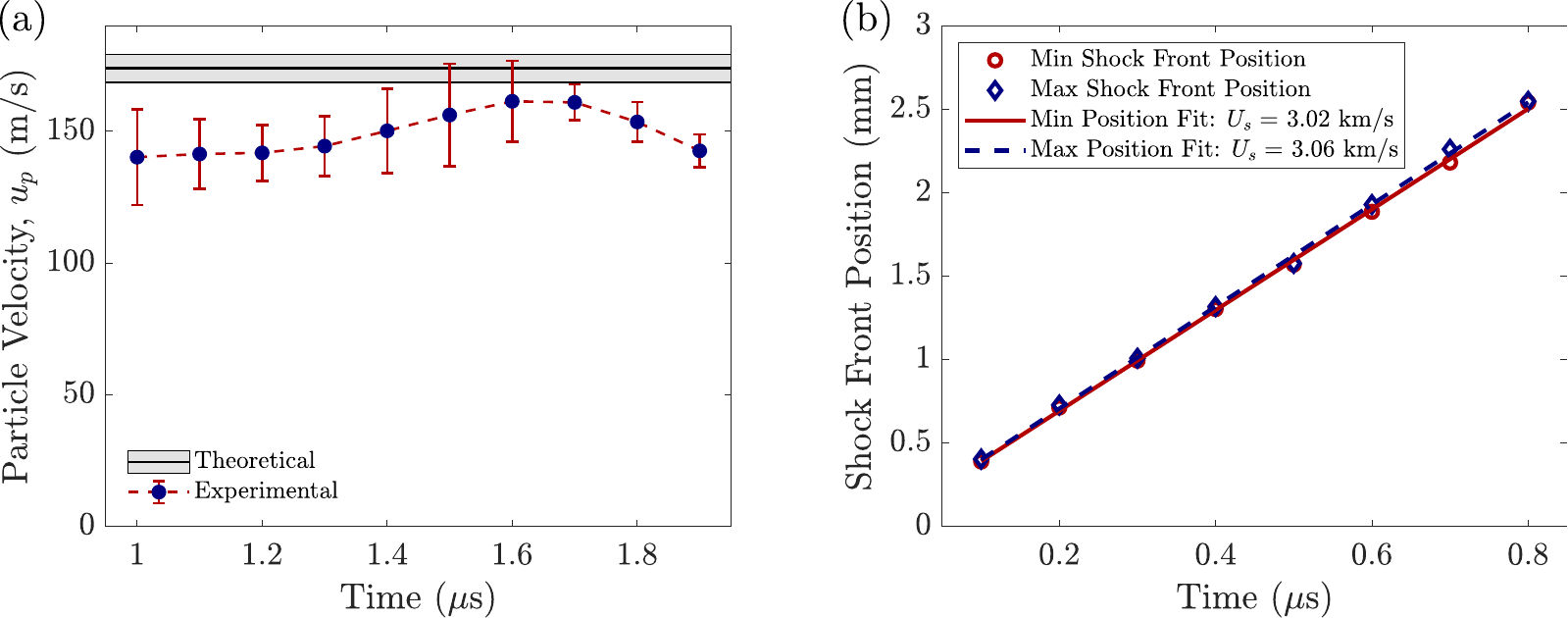}}
            \caption{Velocity measurements for plate impact experiment. Time is shifted such that $t=0$ corresponds with the arrival of the shock front in the field of view. (a) Averaged experimental particle velocity within the area of interest, computed from DIC displacement measurements via central difference. Experimental measurements are compared to the computed theoretical particle velocity, which is calculated through the impedance matching technique based on measured impact velocity and known material properties (see \mbox{\cite{Meyers1994DBOM}}). Experimental error bars include one standard deviation from the mean in addition to $2\%$ error to account for error in the spatial scale. Theoretical error bars account for error in the impact velocity measurements and in the empirical equation of state data. (b) Shock front position measurements, estimated by manually identifying the location where the dark band (indicative of the shock front) begins. The linear fits to the position data provide an estimate of the shock front velocity, which is very close to the theoretically predicted value of $3.10\,$km/s.}

            \label{fig:Velocity Plot}
        \end{figure}

        While the primary focus of the technique and the validation experiments is to perform internal strain measurements, a natural second validation metric is the internal particle velocity. For comparison, the theoretical particle velocity can be predicted through the impedance matching technique as described above. The experimentally measured particle velocity is then computed via central difference of the displacements measured with DIC. \Cref{fig:Velocity Plot}\hyperref[fig:Velocity Plot]{a} shows the average particle velocity in the area of interest for the duration of the experiment, compared with the theoretical particle velocity. The velocity measurements are relatively steady, as is expected for the post-shocked state, and demonstrate reasonable agreement with theory, with an error of approximately $10-30\,$m/s. Interestingly, the particle velocity error percentage coincides closely with that of the strain measurements. This result provides additional confidence in the measurements as shock strain and particle velocity are expected to scale proportionally with one another (\cref{eq:ShockStrain}). It also suggests that the source of error could be a physical effect (as opposed to an optical distortion) such as deviation from the ideal uniaxial strain conditions which are assumed in one dimensional shock theory, as was mentioned previously. Alternatively, it is possible that the impedance matching technique over-predicts the theoretical strain and particle velocity for this experiment, owing to the ubiquitous shock response of PMMA at low pressures \cite{Barker1970PMMA, lacina2018shock, jordan2016shock} and possible variance in PMMA properties depending on the manufacturer. One must also note the importance of the scale calibration used to extract physical displacement measurements based on pixel values. While the strain measurements are insensitive to scale (being a non-dimensionalized quantity), the error in velocity measurements corresponds directly with the error in the calibration scale.

        In addition to the particle velocity measurements, the shock front position is tracked and used to estimate the shock velocity ($U_s$) in the experiment. This is shown in \Cref{fig:Velocity Plot}\hyperref[fig:Velocity Plot]{b}, which indicates a nearly constant shock velocity, and the associated linear fits estimate the value to be between $3.02-3.06\,$ km/s. This is in excellent agreement with the predicted shock velocity of $3.10\,$km/s.

        The present validation experiment demonstrates the feasibility of the technique for application in shock compression of transparent materials. This technique, which is the first to enable accurate, full-field deformation measurements inside shocked materials, shows promise and presents significant opportunities to study the local material response near heterogeneities under dynamic loading conditions \cite{lawlor2024porecollapse}.


\section{Conclusion} \label{sec:Conclusion}
An experimental technique to perform internal DIC has been developed to characterize the large deformation strain fields inside of transparent specimens in the high-strain rate regime with traditional full-scale, dynamic, high-strain rate laboratory experiments. The technique is demonstrated in PMMA for both SHPB and normal plate impact experiments, but it should be possible to implement with many transparent materials and in conjunction with most laboratory dynamic experiments. This represents a critical experimental development enabling investigation of complex phenomena occurring under extreme dynamic loading, such as the deformation and failure of heterogeneous materials at the mesoscale, the local material response under complex loading states, and  the evolution of failure modes which are sensitive to boundary effects.

Validation experiments were performed to determine the feasibility and accuracy of the technique. It was found to be quite accurate ($0.2$\% strain noise, which approaches the noise floor of conventional static DIC) in SHPB experiments during the initial deformation period, though the duration for DIC measurements was shortened by the brittle fracture response of the inner-plane specimens. In the future, this issue could be mitigated for homogeneous samples by adopting the embedded speckle plane patterning technique used elsewhere \cite{Berfield2007Micro,McGhee2023Microcavitation}, though clever solutions would be necessary when using that patterning method for heterogeneous materials. When the current internal DIC technique was applied to plate impact experiments, it was found to be accurate to within about $1\%$ strain and viable up to approximately $0.65\,$GPa impact stress. Several possible error sources are identified, and solutions are presented where applicable. Different material systems, larger specimen size (diameter of the gun barrel), alternative magnifications, improved camera hardware, and better speckle patterning may enable improvement of the technique by mitigating the aforementioned sources of error, thus improving image quality and DIC resolution. Still, the accuracy achieved and the stress regime accessed for the experiments were sufficient to demonstrate the feasibility of the technique in extreme environments, which push the limit of state of the art high-speed cameras.

Future work will focus on extending this technique beyond $0.65\,$GPa impact stress, incorporating modifications to minimize the error, and implementing the technique in various alternative experimental setups and with different transparent materials. Further, the internal DIC technique will be used to study the local shock response of heterogeneous materials. Namely, interest lies in examining porous materials through investigation of the pore collapse phenomenon \cite{Escauriza2020Collapse,Lovinger2024Localization,lawlor2024porecollapse}, and particulate composites through study of the material response and interaction at the interface between hard inclusions and polymer matrix.

\bmhead{Acknowledgements}
The research reported here was supported by the DOE/NNSA (Award No. DE-NA0003957), which is gratefully acknowledged. The support of the Army Research Laboratory (Cooperative Agreement Number W911NF-12-2-0022) for the acquisition of the high-speed camera is acknowledged.

\section*{Declarations}

\subsection*{Competing Interests}
\noindent The authors have no conflicts to disclose.

\subsection*{Author Contributions}
\noindent\textbf{Barry Lawlor:} Conceptualization (lead); Methodology (lead); Investigation (lead); Formal analysis (lead); Visualization (lead); Writing – original draft (lead); Writing – review and editing (equal)\\
\textbf{Vatsa Gandhi:} Conceptualization (supporting); Methodology (supporting); Investigation (supporting); Writing – review and editing (equal) \\
\textbf{Guruswami Ravichandran:} Supervision (lead); Funding acquisition (lead); Conceptualization (supporting); Formal analysis (supporting); Writing – review and editing (equal). \\

\subsection*{Data Availability}
\noindent The data that support the findings of this study are available from the corresponding author upon reasonable request.






\bibliography{References}

\end{document}